\documentclass[aps,pre,onecolumn,notitlepage,showpacs,superscriptaddress,10pt]{revtex4-1}

\usepackage{bm}
\usepackage{graphicx}
\usepackage{latexsym}
\usepackage{amsmath}
\usepackage{hyperref}
\usepackage{pxfonts}
\usepackage{cancel}
\usepackage{bbding}
\usepackage{subfig}
\usepackage{float}
\usepackage{subfloat}
\usepackage{caption}
\usepackage[section]{placeins}
\usepackage{csquotes}
\usepackage{relsize}
\usepackage{enumitem}
\DeclareGraphicsExtensions{.pdf,.png,.jpg}

\LetLtxMacro{\OldSqrt}{\sqrt}
\newcommand{\ClosedSqrt}[1][\hphantom{3}]{\def\DHLindex{#1}\mathpalette\DHLhksqrt}
\makeatletter
    \newcommand*\bold@name{bold}
    \def\DHLhksqrt#1#2{%
        \setbox0=\hbox{$#1\OldSqrt{#2\,}$}\dimen0=\ht0\relax%
        \advance\dimen0-0.2\ht0\relax
        \setbox2=\hbox{\vrule height\ht0 depth -\dimen0}%
        {\hbox{$#1\expandafter\OldSqrt\expandafter[\DHLindex]{#2\,}$}
        \lower\ifx\math@version\bold@name0.6pt\else0.4pt\fi\box2}
    }
    \renewcommand*{\sqrt}[2][\ ]{\ClosedSqrt[\leftroot{-2}\uproot{1}#1]{#2}\kern0.1em} 
\makeatother

\begin{document}

\title{Laser-driven plasma acceleration in a regime of strong-mismatch between the incident laser envelope and the nonlinear plasma response}

\author{A. A. Sahai}
\affiliation{John Adams Institute for Accelerator Science, Department of Physics, Imperial College London, SW7 2AZ, United Kingdom}
\email[corresponding author: ~]{a.sahai@imperial.ac.uk}
\author{K. Poder, J. C. Wood, J. M. Cole, N. C. Lopes, S. P. D. Mangles, Z. Najmudin}
\affiliation{John Adams Institute for Accelerator Science, Department of Physics, Imperial College London, SW7 2AZ, United Kingdom}
\date{8 March 2017}

\maketitle
\vspace{-0.5in}

\section*{Abstract}\label{sec:abstract}

We explore a regime of laser-driven plasma acceleration of electrons where the radial envelope of the laser-pulse incident at the plasma entrance is strongly mismatched to the nonlinear plasma electron response excited by it. This regime has been experimentally studied with the \textsc{gemini} laser using f/40 focusing optics in August 2015 and f/20 in 2008. The physical mechanisms and the scaling laws of electron acceleration achievable in a laser-plasma accelerator have been studied in the radially matched laser regime and thus are not accurate in the strongly mismatched regime explored here. In this work, we show that a novel adjusted-$a_0$ model applicable over a specific range of densities where the laser enters the state of a strong optical shock, describes the mismatched regime. Beside several novel aspects of laser-plasma interaction dynamics relating to an elongating bubble shape and the corresponding self-injection mechanism, importantly we find that in this strongly mismatched regime when the laser pulse transforms into an optical shock it is possible to achieve beam-energies that significantly exceed the incident intensity matched regime scaling laws. 

\section{Introduction}\label{sec:introduction}
Plasma accelerators \cite{tajima-dawson} rely on laser-excited plasma electron response for keeping a laser pulse ($\omega_0$,$\lambda_0$,$v_g$) continuously focussed to high-intensities over a distance which is many times its Raleigh length, $\rm{Z_R}$ to sustain a regime of high-amplitude plasma wave [$\omega_{pe}=n_0e^2\epsilon_0^{-1}m_e^{-1}$,$k_{pe}=\omega_{pe}/v_{\phi}$] ($v_{\phi}$, the plasma-wave phase velocity nearly equal to $v_g$). The self-guided regime \cite{Max-1974}\cite{Sun-1987}\cite{Sprangle-1987} utilizes relativistic and ponderomotive channeling effects to preserve the laser intensity over the acceleration length, thus overcoming the external guiding requirements. But to enable self-guiding, the laser power has to be higher than the critical power, $\rm{P_c}$ which depends inversely on the plasma density, $n_0$. Thus laser power dictates the densities that a self-guided laser-plasma accelerators can operate at to obtain certain electron beam energies. The analysis presented here is in the self-guided regime. 

Typically, for a laser pulse to self-guide its ponderomotive force has to cavitate electrons, where the plasma electrons in its path are nearly completely expelled. Whereas this leads to a radial profile of the refractive index \cite{Esarey-1996}, $\eta(r,z)$ which is favorable for guiding the laser pulse, it also drives a plasma wave which is in the non-linear regime, $\delta n_e(r,z)/n_0 \equiv (n_e(r,z)-n_0)/n_0 > 1$. The results presented here are in the non-linear cavitated or the bubble regime of plasma wave \cite{Pukhov-bubble-2002}\cite{mangles-2004}\cite{Pukhov-2005}.



\begin{figure}[!htb]
   \includegraphics*[width=\columnwidth]{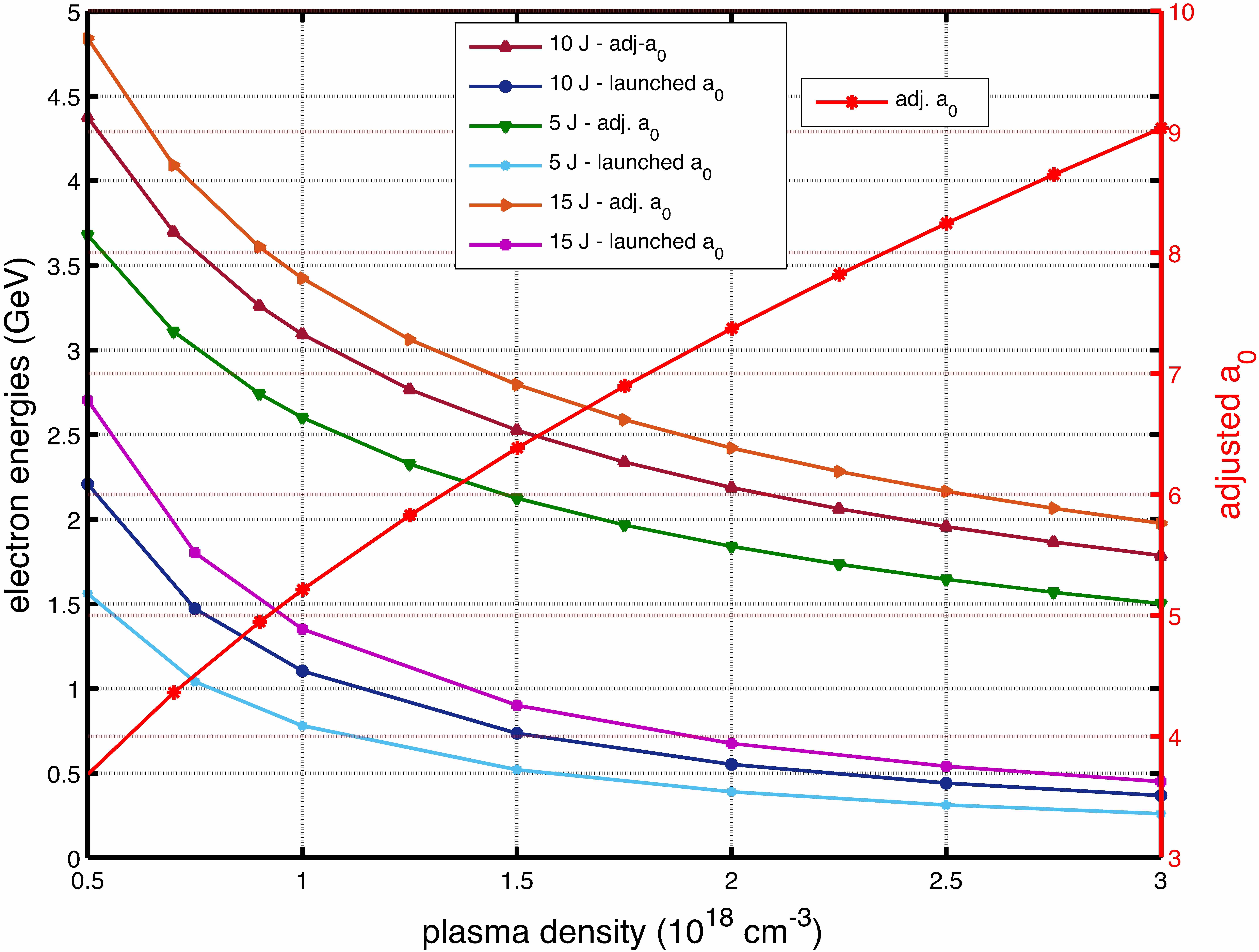}
   \caption{ Electron beam energy in the bubble regime in the matched regime \cite{Lu_PRSTAB_2007} for $w_0$ corresponding to \textsc{gemini} f/40 optics for laser energies $\rm{\mathcal{E}_L}$=5,10 \& 15J. Corresponding beam energies for $a_0$ values {\it adjusted} for the matched condition, $a_0(\rm{adj})$ (shown for Table.\ref{table:Gemini-f40-parameters} parameters).}
   \label{Fig1:e-beam-energy-scaling}
\end{figure}

In the bubble regime the laser excited non-linear plasma response (electron-ion charge separation) equilibrates with the laser ponderomotive force such that the laser waist-size is \cite{Pukhov-2005}\cite{Lu_PRSTAB_2007}, 
\begin{equation}
\begin{aligned}
w_0 \simeq 2 \sqrt{a_0} ~ \frac{c}{\omega_{pe}} \equiv R_{\rm{bubble}}
\end{aligned}
\label{eq:radial-matched-condition}
\end{equation}
\noindent where, $w_0$ is the laser focal spot-size, $a_0=\rm{max}[e\vec{A}/m_ec^2]=8.55\times10^{-10}\lambda_0[\mu m]\sqrt{I_0[\rm{W/cm^2}]}$ the laser strength parameter, $\vec{A}$ the laser vector potential, $I_0$ is the peak intensity and $\lambda_0$ the laser wavelength. 

In the bubble regime when a laser pulse is coupled into the plasma with a waist-size $w_0$ related to its strength parameter $a_0$ by eq.\ref{eq:radial-matched-condition}, its radial envelope oscillations are minimized due to the equilibrium forced by the initial conditions, this is the {\it matched condition}. In the analysis of the laser-plasma conditions for the {\it matched regime} it is can be shown to be the {\it optimal regime} for electron beam energy gain. The scaling law of the electron beam energy gain in the matched regime based upon 3D PIC simulations is \cite{Lu_PRSTAB_2007}:
\begin{equation}
\begin{aligned}
\Delta \mathcal{E} ~ [m_ec^2] \simeq \frac{2}{3} ~ a_0 ~ \left(\frac{n_c}{n_0}\right)
\end{aligned}
\label{eq:radial-matched-condition-energy}
\end{equation}
\noindent The laser strength parameter in eq.\ref{eq:radial-matched-condition-energy}, is the launched $a_0$ at the plasma entrance. The value of $a_0$ is known to vary over the acceleration length due to several non-linear laser-plasma interactions effects such as self-compression, pump depletion and head energy-loss of the laser pulse. 

\begin{table}[!htb]
\setlength{\tabcolsep}{20pt}
\begin{center}
\captionsetup{belowskip=5pt,aboveskip=2pt}
\caption{\textsc{gemini} laser \& $e^-$ beam parameters}
\label{table:Gemini-f40-parameters}
\begin{tabular}{ l  l }
\\ [-3.0ex]\hline 
\hline \\[-2.4ex]
 	$\rm{\mathcal{E}_L}$, FWHM-$\mathlarger{\mathlarger{\tau_p}}$ 				& $\simeq$ 10 J, 49 fs \\
	$P_L$																		& $\simeq$ 200 TW \\
  	$w_{0-y}$ (y-axis) 				 											& 37.4 $\mu m$  \\
	$w_{0-z}$ (z-axis) 				 											& 44.2 $\mu m$  \\
	$\rm{peak} ~ a_0$ (coupled)													& $\simeq 1.9$ \\
 	$\Delta\mathcal{E}_{\rm{peak}}$												& 2.2 GeV \\
	$n_0$ (peak energy)			    											& $2-3 \times 10^{18} ~ \rm{cm^{-3}}$ \\
	$P_c$																		& $18.3 - 12.2$ TW \\
\hline	 
\end{tabular}
\end{center}
\end{table}%

The estimated energies over a range of densities using eq.\ref{eq:radial-matched-condition-energy} for the matched regime with \textsc{gemini} f/40 parameters (summarized in Table.\ref{table:Gemini-f40-parameters}) are shown in Fig.\ref{Fig1:e-beam-energy-scaling}. It is quite evident from Fig.\ref{Fig1:e-beam-energy-scaling} that the peak energies predicted by eq.\ref{eq:radial-matched-condition-energy} for the \textsc{gemini} f/40 experiments with $n_0 = 1.5-3 \times 10^{18} ~ \rm{cm^{-3}}$ are $\leq \rm{1GeV}$. Thus, there is a striking dis-agreement between the predictions of the 3D simulation based models of \cite{Lu_PRSTAB_2007} and the experimental observations. 

\section{Strongly mismatched regime \& the adjusted-$a_0$ model}

To understand this significant diagreement between the predicted energies and the experimentally obtained electron beam energies, it is firstly important to note that the \textsc{gemini} f/40 results are in a {\it strongly mismatched regime}. The reason for this is evident from eq.\ref{eq:radial-matched-condition}, whereas the matched $w_0$ for this interaction is $10.3\mu m$ the launched elliptical laser pulse has its smaller waist-size of $37.4 ~\mu m$.

The {\it strongly mismatched regime} of laser-plasma acceleration is considered sub-optimal because the laser radial envelope oscillates around the equilibrium matched spot-size which is expected to continuously change the plasma wave dimensions. This is expected to result in the optimal wave phase for acceleration and focusing to continuously change its position relative to the beam. In comparison to the matched regime, here the beam is loaded in sub-optimal phases. It can thus be argued that the effective acceleration field and acceleration distance, are as a result, sub-optimal. This conventional understanding of the laser-plasma acceleration process developed with the matched regime as the reference, seems to be entirely inapplicable to the experimental results obtained in the {\it strongly mismatched regime}. 

It is however still possible to argue using the conventional understanding by applying an {\it adjusted-$a_0$ model}. This model assumes that the entire laser pulse energy launched at the entrance of the plasma is coupled into the plasma and is then squeezed to the matched spot-size corresponding to the launched $a_0$. This will therefore increase the $a_0$ by the factor $w_0(\text{launched}) / w_0(\text{matched})$ upon the culmination of the squeezing process for a radially symmetric focal spot. The energy gain in this {\it adjusted-$a_0$ model} can thus be written as in eq.\ref{eq:energy-strong-mismatch} where subscript `l' are for the launched quantities and `m' for matched ones. F is the F-number of the focal spot, $F=\frac{\pi w_0}{2\lambda_0}$. It is related to the F-number of the focusing optics ($F=f/D$, $f$ being the focal length and $D$ being the aperture of the lens).

\begin{equation}
\begin{aligned}
\Delta \mathcal{E}_{\text{adj.}}[m_ec^2] & = \frac{2\pi}{3} ~ \sqrt{a_{0-l}} ~ \left(\frac{w_{0-l}}{\lambda_0}\right) ~ \sqrt{\frac{n_c}{n_0}} \\
& \simeq 2.5 ~ \sqrt{a_{0-l}} ~ \sqrt{\frac{n_c}{n_0}} ~ \text{F}_l  \\
{\rm circ. : ~} a_0 (\text{adj.}) & = a_{0-l} ~ \left(\frac{w_{0-l}}{w_{0-m}}\right) \\
{\rm ellip. : ~} a_0 (\text{adj.}) & = a_{0-l} ~ \sqrt{\frac{w_{01-l} ~ w_{02-l}}{w_{0-m}^2}}
\end{aligned}
\label{eq:energy-strong-mismatch}
\end{equation}

We can then calculate the expected electron energies using $a_0(\rm{adj.})$ and as shown in Fig.\ref{Fig1:e-beam-energy-scaling} and eq.\ref{eq:energy-mismatch-f40}, a good match to experimental electron energies is obtained. For example, at $n_0 = 2 \times 10^{18} ~ \rm{cm^{-3}}$ the matched spot-size, $w_0(\text{matched}) = 10.2 ~ \mu m$.

\begin{equation}
\begin{aligned}
\noindent \rm{f/40} ~ optics ~ \rm{at} ~  2 \times 10^{18} \rm{cm^{-3}}, & ~ \mathcal{E}_L = 10\rm{J}, a_0 \simeq 1.9 \\
\rm{\Delta \mathcal{E} ~ eq.\ref{eq:radial-matched-condition-energy} :} & < \rm{1 ~ GeV} \\
\rm{peak ~ expt. ~ \Delta \mathcal{E} :} & ~ \rm{2.2 ~ GeV} \\
\rm{\Delta \mathcal{E} ~ [a_0(adj) = 7.4] :} & ~ \rm{2.2 ~ GeV}
\end{aligned}
\label{eq:energy-mismatch-f40}
\end{equation}
Here, because the \textsc{gemini} f/40 focal spot is elliptical we have used $a_0(\text{adj.}) = a_0 ~ (w_0[y]w_0[z](\text{launched})/w_0^2(\text{matched}))^{0.5}$.

We find that the peak plasma fields in this regime are of the order of $a_0(\rm{adj.}) \times m_ec\omega_{pe}e^{-1}$. At $2\times10^{18} \rm{cm^{-3}}$, $m_ec\omega_{pe}e^{-1} \equiv \rm{E_{wb}} = 136 ~ GVm^{-1}$ and the predicted peak fields are $a_0(\rm{adj.}) \times m_ec\omega_{pe}e^{-1} = 1006 ~ GVm^{-1}$. For the laser parameters in Table.\ref{table:Gemini-f40-parameters}, the value of average accelerating plasma field from \cite{Lu_PRSTAB_2007} is $\rm E_{plasma}(a_0=1.9) = 0.5 \sqrt{a_0} ~ m_ec\omega_{pe}e^{-1} = 93.7 ~ GVm^{-1}$ or with $a_0(\rm{adj.})$ it is $\rm E_{plasma}(a_0[adj.]) = 185 ~ GVm^{-1}$. The {\it adjusted-$a_0$ model} predicts quite correctly even in this case - in experiments we have peak beam energies of 2.2 GeV in 20mm. But, from the simulations the injection of the high-energy bunch occurs only around 24 ps. This gives us an average acceleration gradient of $\simeq 175 ~ \rm GVm^{-1}$.

Not surprisingly, an agreement using these arguments is also obtained for \textsc{gemini} f/20 data \cite{Kneip-2009} where the matched $w_0$ is $8.95 ~ \mu m$ at $5.5 \times 10^{18} ~ \rm{cm^{-3}}$ with $a_0=3.9$ whereas the launched $w_0=19\mu m$. The energy expected from the matched regime formula in eq.\ref{eq:radial-matched-condition-energy} is 510 MeV. Using the adjusted $a_0$-model $a_0 \rm{(adj.)}=8.3$ with expected beam energy of $957 \rm{MeV}$.

\begin{equation}
\begin{aligned}
\noindent \rm{f/20} ~ optics ~ \rm{at} ~  5.5 \times 10^{18} \rm{cm^{-3}}, & ~ \mathcal{E}_L = 10\rm{J}, a_0=3.9 \\
\rm{\Delta \mathcal{E} ~ eq.\ref{eq:radial-matched-condition-energy} :} & < \rm{510 ~ MeV} \\
\rm{peak ~ expt. ~ \Delta \mathcal{E} :} & ~ \rm{800 ~ MeV} \\
\rm{\Delta \mathcal{E} ~ [a_0(adj) = 8.3]:} & ~ \rm{957 ~ MeV}
\end{aligned}
\label{eq:energy-mismatch-f20}
\end{equation}

It will be shown using PIC simulations that this argument is not arbitrary because the assumption of laser energy squeezing down to the matched spot-size and the laser waist subsequently remaining close to the launched $a_0$ matched size holds (see Fig.\ref{Fig3:combined-evolution-f40}). However, there are a couple of major questions that go against generalizing this argument:
\begin{enumerate}[nolistsep,,label=(\roman*)]
\item Why does this energy scaling only work optimally for densities in the range $1.5 - 3 \times 10^{18} ~ \rm{cm^{-3}}$ for the f/40 focusing ?
\item Why is there a similar optimal density in the range $5-7 \times 10^{18} ~ \rm{cm^{-3}}$ for the results with \textsc{gemini} f/20 focusing optics data \cite{Kneip-2009} ? 
\end{enumerate}
Is it a result of the squeezing process only working over certain densities ? 

It is clear from the above contradictions that even though the {\it adjusted-$a_0$ model} is quite applicable over a density range, it cannot be universally applied. We have investigated the reasons behind this and have been able  to check using simulations that the envelope squeezing effect occurs for a broad range of densities in comparison to the densities over which the experiments produce peak electron energies. Therefore, there must be another reason for the {\it adjusted-$a_0$ model} working only over a small range of densities. 

We observe certain interesting effects in the simulations that persist only over a certain range of densities. As the laser focusses down to the matched focal spot, the non-linear plasma response to the now much higher laser intensity tries to equilibrate it to a larger spot-size but during this expansion the laser loses a large part of the head of its longitudinal envelope and enters an {\it optical shock} regime \cite{optical-shock}. In the optical shock regime the longitudinal ponderomotive force far exceeds the radial ponderomotive force resulting in the lengthening of the bubble, while the radial excursions are inhibited (see Fig.\ref{Fig5:ponderomotive-force-evolution}). 

Therefore, the acceleration dynamics in the strongly mismatched regime is dictated by the laser envelope dynamics (in line with expectations due to the strong radial mismatch initial conditions) triggering the onset of {\it optical shock} state of the laser. We present more detailed analysis of this in the simulations section.

\begin{figure}[!hb]
   \includegraphics*[width=4.5in]{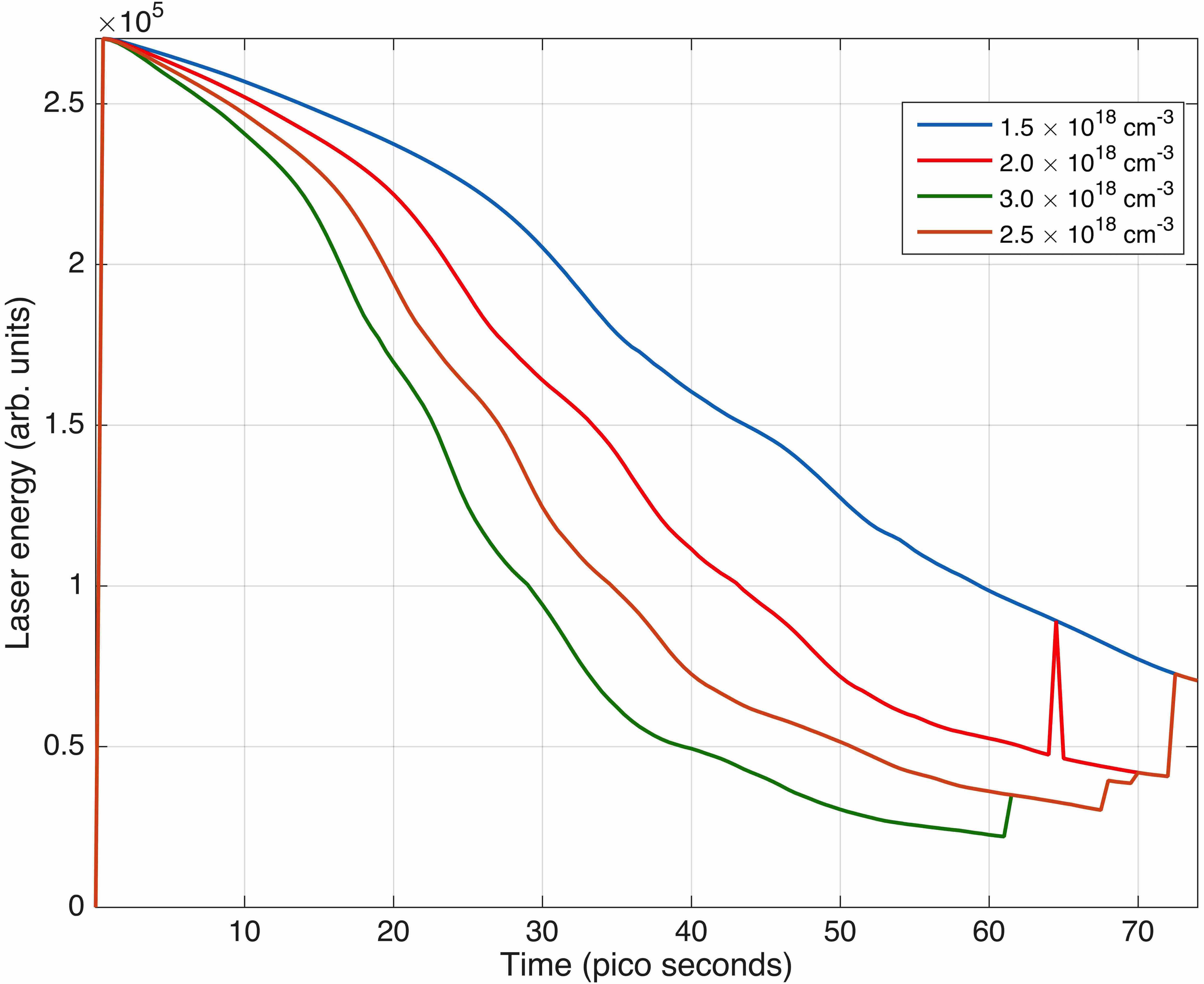}
   \caption{ Laser energy evolution with propagation distance for different densities from 2D PIC simulations}
   \label{Fig2:laser-energy-evolution}
\end{figure}

\begin{figure}[!ht]
   \includegraphics*[width=4.5in]{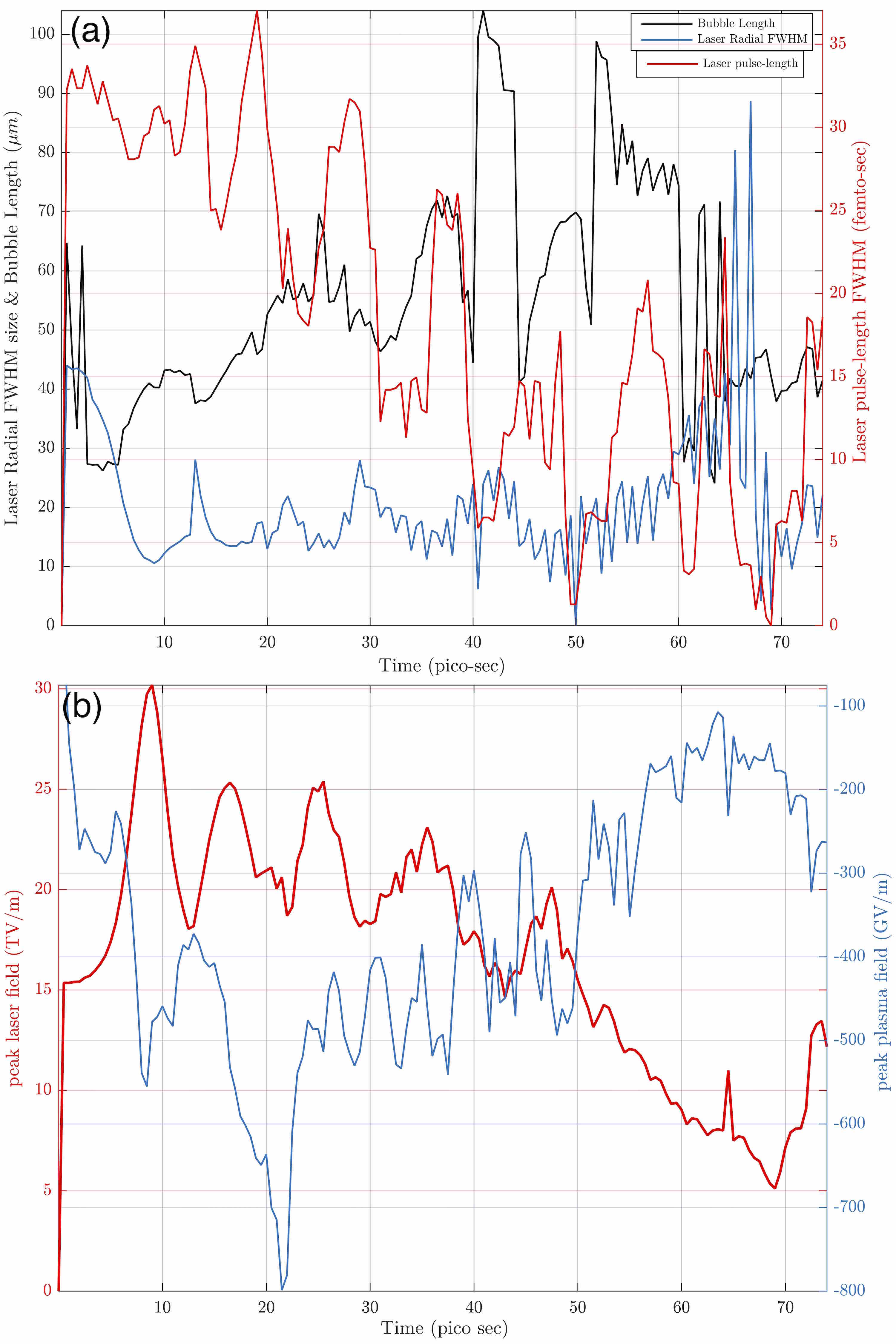}
   \caption{ 2D PIC simulations showing the evolution of peak laser and peak electron acceleration plasma field with propagation distance using the f/40 focusing optics parameters for $\rm{\mathcal{E}_L}$ = 10J. }
   \label{Fig3:combined-evolution-f40}
\end{figure}

\subsection{Experimental considerations for a large focal-spot}

From an experimental perspective, achieving a good quality Gaussian focal spot (measured with $\rm M^2$-number) is practically easier for a larger spot-size. In theory, even though a matched spot-size may be ideal for generating a stable plasma-wave with slowly evolving dimensions, in practice focusing to the matched spot-size for densities in $10^{18}$ might lead to a non-uniform Gaussian spot with multiple hot-spots. 

A radially non-uniform focal-spot affects the transverse characteristics of the plasma wave, leading to non-optimal acceleration and focusing field profiles. So, there is definitely an experimental justification to the observation of larger focal spots leading to higher energies. The laser radial profile at the plasma exit may be utilized to investigate the transverse characteristics of the plasma wave. The presence of multiple hot-spots in the incident focal spot may be inferred by observing the exit mode of the laser, as seen in Fig.3(c)-(h) in \cite{Kneip-2009}.

Note that, if the squeezing process can be confirmed experimentally, this would be an {\it optical plasma lens}. Because, as we show below that the energy loss of the laser during the compression phase is negligible. This is quite similar but operates based on different physical mechanisms than a {\it beam plasma lens} \cite{chen-lens-1986}.

\section{Analysis based upon PIC Simulations}\label{sec:simulations}

In this section we present results from 2D Particle-In-Cell (PIC) simulations carried out using \textsc{epoch}. We use a moving simulation box which tracks the laser pulse (Table \ref{table:Gemini-f40-parameters} parameters with $a_0$ boosted by a factor of 2 to compensate for peak self-focussed $a_0$ in 2D simulations not reaching $a_0(\text{adj.})$) at a velocity equal to the laser group velocity in the unperturbed plasma. The box uses a cartesian grid with 25 cells per laser wavelength ($\lambda_0$) in the longitudinal direction and 10 cells per laser wavelength in the transverse with 4 particles per cell. Absorbing boundary conditions are used for both the fields and particles. The laser pulse is incident from the left boundary and focusses to the smaller focal spot-size (of the elliptical \textsc{gemini} f/40 focal spot) onto homogeneous plasma before the box starts moving. The plasma is setup to have a 500$\mu m$ linear density gradient before the homogeneous plasma, to mimic the density profiles in a gas-cell.

In Fig.\ref{Fig2:laser-energy-evolution} the evolution of laser energy with time is shown for different densities. It is well known that the laser energy depletion goes as $n_e/n_c$, which is what we observe here. However, the laser energy evolution does not have any correlation with the electron beam energies or specific signature about laser-plasma interaction process in the mismatched regime. Interestingly, the energy loss during the ``squeezing"-phase is relatively small, allowing the non-linear regime to be useful as an efficient {\it optical plasma lens} and the {\it adjusted-$a_0$} model to be valid. 

In Fig.\ref{Fig3:combined-evolution-f40}, the evolution of the laser-plasma dimensions and laser-plasma fields are plotted for $n_0 = 2\times10^{18} \rm{cm^{-3}}$. There are several interesting laser-plasma effects that can be inferred from the evolution of the laser-plasma dimensions in Fig.\ref{Fig3:combined-evolution-f40}(a). 

\begin{enumerate}[nolistsep,label=(\roman*)]
\item The laser pulse intensity-FWHM waist size (in blue) launched at 44$\mu m$ is squeezed down to the matched spot-size of $\simeq 10\mu m$ in 10 ps ($\simeq$3mm). This process of the squeezing down of the initial focal spot to nearly the matched spot-size of the launched $a_0$ occurs for nearly all the simulations over a wide range of densities (data not presented here).
\item The laser pulse {\it radial envelope} oscillates due to the strong initial mismatch. However, the spot-size remains close to the matched spot-size corresponding to the launched $a_0$. The maximum radial excursions are of nearly half the launched spot-size however importantly these are all {\it precursors} to the triggering of optical shock state. This explains the agreement of the experiments in the {\it strongly mismatched regime} to the {\it adjusted-$a_0$ model}.
\item The laser pulse {\it longitudinal or temporal envelope} undergoes a few {\it catastrophic} collapses. This is inferred by observing the intensity-FWHM time duration of the laser pulse (in red) over time. There are about 4 such events around 14ps, 21ps, 29ps and 40ps. The collapse of the laser pulse leads to the formation of an {\it optical shock}, shown in Fig.\ref{Fig4:optical-shock-formation}.
\item The effect of this collapse is observed in a corresponding increase in the {\it bubble length}. The length of the bubble starts at nearly the same radius as laser radial envelope but does not reduce as fast as the laser radial FWHM. The length increases while it radius remains nearly constant, as seen by comparing the blue and the black curve. This leads to the optical shock driven excitation of {\it asymmetric elongated bubble}.
\item The triggering of {\it optical shock} formation and the excitation of an {\it asymmetric bubble} corresponds directly to the injection of electrons in the back of the {\it elongating} bubble. 
\end{enumerate}

The laser-plasma interactions effects driving the acceleration are also analyzed by observing the evolution of the laser and plasma fields in Fig.\ref{Fig3:combined-evolution-f40}(b). It is quite clear that the peak plasma field occurs when the laser pulse temporal intensity-FWHM undergoes a sudden collapse. We observe that this dip in temporal intensity-FWHM corresponds to the development of an optical shock. The plasma field is at its peak of $\simeq -800 ~ \text{GV/m}$ at around 21 ps. 

The highest energy bunches are injected around 24ps when the bubble is rapidly elongating in response to the largely dis-balanced and increasing longitudinal ponderomotive force due to excitation of an optical shock, a condition represented in eq.\ref{eq:dis-balanced-ponderomotive-force}, where $\mathcal{E}_{quiv}^e(x,r)$ is the energy of quiver motion in the laser field.

\begin{equation}
\begin{aligned}
& \mathcal{E}_{quiv}^e(x,r) \propto I(x,r)\lambda_0^2(x,r) \\
\text{Spherical bubble :} & \nabla_x \mathcal{E}_{quiv}^e(x,r) \simeq \nabla_r \mathcal{E}_{quiv}^e(x,r)  \\
\text{Elongated bubble :} & \nabla_x \mathcal{E}_{quiv}^e(x,r) \gg \nabla_r \mathcal{E}_{quiv}^e(x,r)
\end{aligned}
\label{eq:dis-balanced-ponderomotive-force}
\end{equation}

At 21 ps, the laser energy and the peak laser field are still high enough to cause the strongest dis-balance between the radial and longitudinal forces of the laser (see Fig.\ref{Fig5:ponderomotive-force-evolution}) during its evolution.

\subsection{Optical shock excitation}

We analyze the {\it triggering} of an optical shock excitation by the evolving dynamics of the laser envelope in the strongly mismatched regime. In this regime, due to the oscillations of the laser radial envelope each of the envelope-squeeze event driving the laser radial envelope towards the matched spot-size causes a large surge in the laser electric field. We show that each of the surge in the laser electric fields triggers an {\it optical shock excitation} event.

\begin{figure}[!htb]
   \includegraphics*[width=\columnwidth]{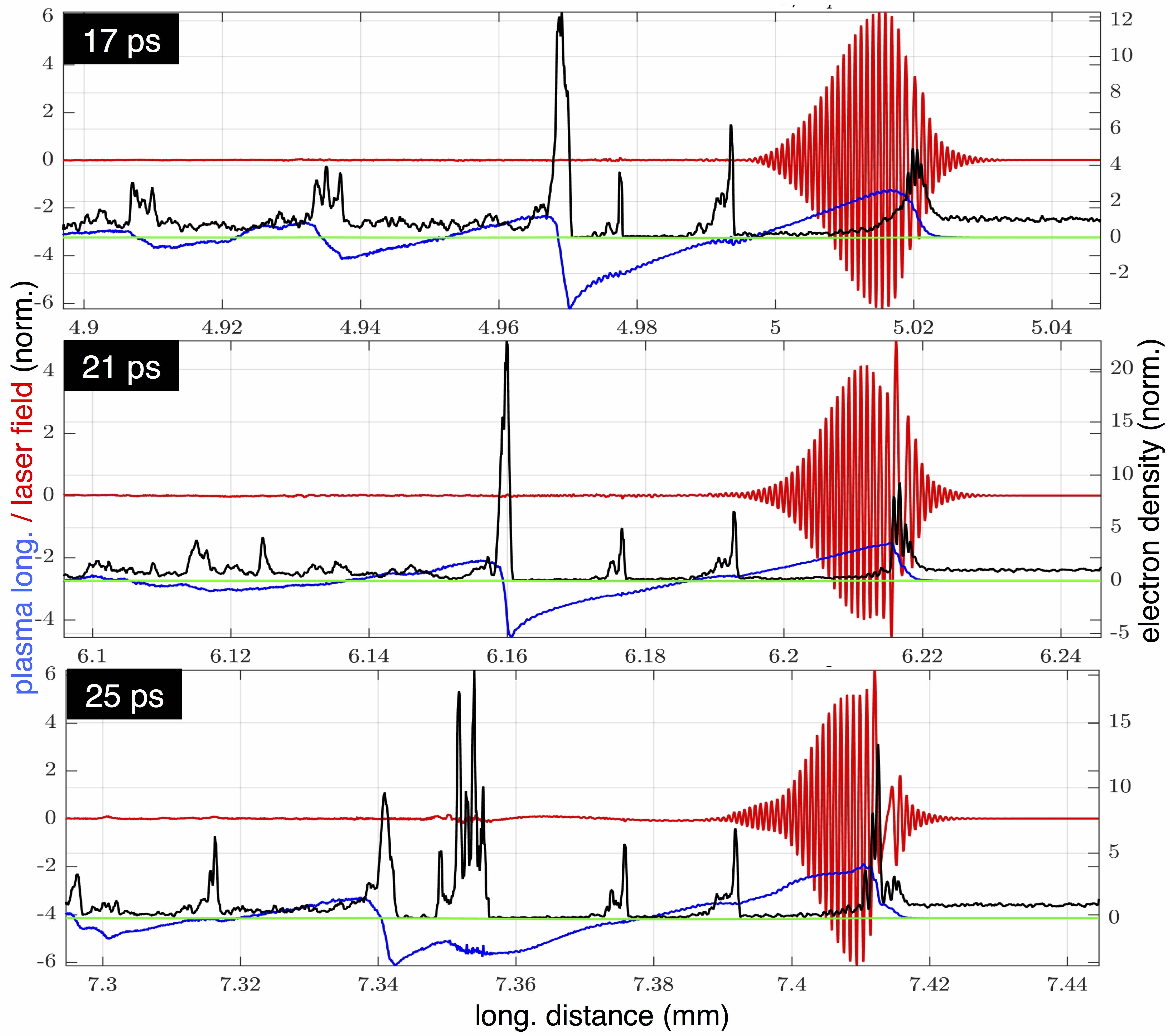}
   \caption{On-axis line-out from 2D PIC simulations showing the formation of the optical shock in the strongly mismatched regime. The red curve is the laser transverse electric field. The plasma longitudinal field (blue curve) can be used to infer the bubble length, which is seen undergoing an increase. The green line is the zero line for the axis on the right-hand side, it sets the zero level for the density (in black) and the plasma longitudinal field.}
   \label{Fig4:optical-shock-formation}
\end{figure}

\begin{figure}
   \includegraphics*[width=\columnwidth]{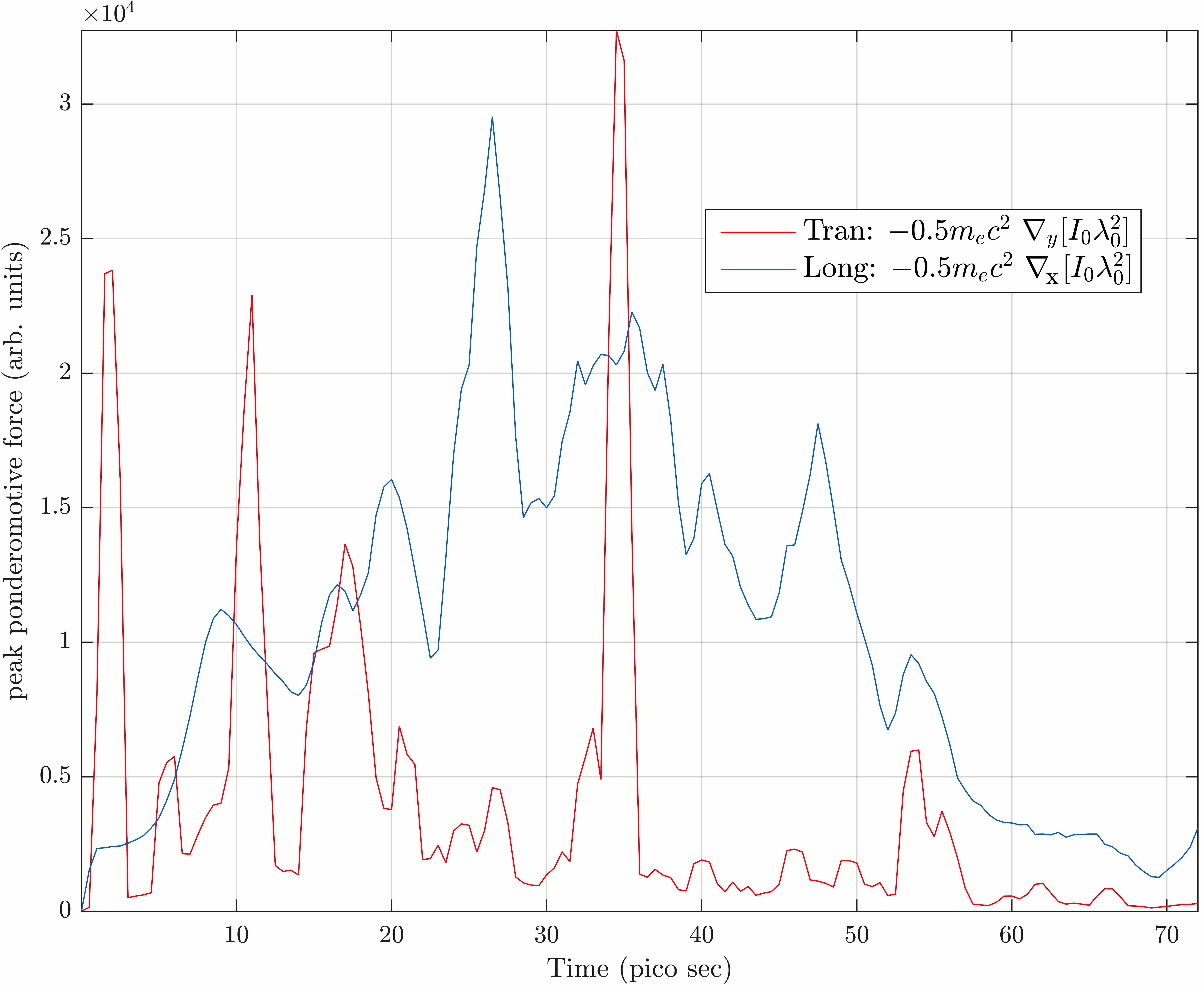}
   \caption{Evolution of the peak ponderomotive force of the laser. The blue curve shows the peak longitudinal ponderomotive force along the axis of the laser and the red curve shows the peak transverse ponderomotive force where the laser field is maximum.}
   \label{Fig5:ponderomotive-force-evolution}
\end{figure}

The third-order perturbative expansion based relation for the laser pulse group velocity \cite{sahai-thesis-2015} in a quasi-static plasma wake with local plasma $\delta n(\xi,r)/n_0$ and laser $a(\xi,r)$ parameters (where, $\xi=c\beta_{g0}t-z$ is a co-moving coordinate to the propagating laser) is in eq.\ref{eq:third-order-dispersion-velocity}. We can use this equation to treat spatially-localized laser-plasma interaction at each instant of the simulation, because it handles group velocity $\beta_g(\xi,r)$ at each point in space in the co-moving coordinate.

\begin{equation}
\begin{aligned}
\beta_g = \beta_{\phi-las}^{-1} \simeq ~ & \beta_{g0} ~ \left[ 1 + \frac{1}{2\gamma_{g0}^2\beta_{g0}^2} \left(\langle a_{\perp} \rangle^2 - \frac{\delta n}{n_0}\right) \right] \\
& \beta_{g0} = \sqrt{1 - \frac{\omega_{pe}^2}{\omega_0^2} } , \gamma_{g0} = \frac{\omega_0}{\omega_{pe}} 
\end{aligned}
\label{eq:third-order-dispersion-velocity}
\end{equation}

This relation in eq.\ref{eq:third-order-dispersion-velocity} can also be used to estimate a {\it locally zero group velocity} condition shown in eq.\ref{eq:zero-local-group-velocity}. 

\begin{equation}
\begin{aligned}
& \beta_g(\xi,r) = 0 \\
& \frac{1}{2}\left(\frac{\delta n}{n_0} - \langle a_{\perp} \rangle^2\right) \simeq \frac{n_c}{n_0}
\end{aligned}
\label{eq:zero-local-group-velocity}
\end{equation}

Though the local envelope velocity being zero might be a hypothetical condition because in this work the typical $n_c/n_0 \simeq 30$, it does demonstrate that the group velocity of a pulse in different parts of the wake can be significantly different. It is quite evident that if $\delta n(\xi,r) \rightarrow n_c$ when $\langle a_{\perp}(\xi,r) \rangle^2\rightarrow 0$, then the local group velocity, $\beta_g(\xi,r)\rightarrow 0$. This implies that a part of the envelope can be slowed down much more in comparison to the rest of the pulse and thus  a part of the envelope can be lost leading to {\it slicing} of the laser into two distinct pulse under the conditions above. It should also be noted that the $\beta_g(\xi,r)\rightarrow 0$ implies $\beta_{\phi-las}(\xi,r)\rightarrow\infty$ which means $\lambda_{las}(\xi,r)\rightarrow \infty$. So, increasing wavelength in a local region implies a local reduction in the group velocity.

The time evolution of the on-axis laser field from PIC simulations is shown for one such event in Fig.\ref{Fig4:optical-shock-formation} which corresponds to the formation of an optical shock at 21ps. 

To understand the dynamics better, we try to simultaneously observe the radial intensity-FWHM in Fig.\ref{Fig3:combined-evolution-f40}(a) and the laser field in Fig.\ref{Fig3:combined-evolution-f40}(b), while observing the on-axis laser dynamics in Fig.\ref{Fig4:optical-shock-formation}.

Now we concentrate on the laser-plasma interaction dynamics in the front of the bubble in Fig.\ref{Fig4:optical-shock-formation}. It can be seen that at 17ps the wavelength in the front of the wake, a region where there is max-$\delta n/n_0$ (where the longitudinal ponderomotive force is the highest), begins to lengthen. This time also corresponds to a large surge in the laser electric field and thus the ponderomotive force is rapidly increasing. This also leads to an increase in the max-$\delta n/n_0$ at the laser head. At 21ps, in the region of max-$\delta n/n_0$ the wavelength has significantly stretched. This corresponds to a rapid reduction in the local group velocity. At 25ps, we observe that the laser envelope is broken into two distinct regions separated by a long wavelength low group velocity cycle. These laser cycles of long-wavelength low group-velocity lead to the head of the laser pulse detaching from it and triggering the laser into an optical shock state. This triggered {\it slicing} is also the time where the laser longitudinal ponderomotive force grows into a large dis-balance with the radial ponderomotive force, as seen in Fig.\ref{Fig5:ponderomotive-force-evolution}.

The large dis-balance between the longitudinal and the radial ponderomotive force is seen to have a direct effect on the dis-balance between the length and the radial envelope of the laser, by comparing Fig.\ref{Fig5:ponderomotive-force-evolution} and Fig.\ref{Fig3:combined-evolution-f40}(a). The bubble length is seen to grow much more than the bubble radius. The large longitudinal ponderomotive force also has a direct effect on the peak of the longitudinal plasma field as seen in Fig.\ref{Fig3:combined-evolution-f40}(b). 

We find that this elongation following an optical shock trigger is the main reason for the self-injection of large amounts of charge in the strongly mismatched regime. Because the injection of charge occurs when the laser is in the state of an optical shock, the injected charge experiences the peak plasma field and accelerates to peak energies in about a centimeter. The elongated bubble also has a longer de-phasing length, thus allowing for slower de-phasing, while it lasts in the elongated state.

\subsection{Beam properties from optical shock injection}

\begin{figure}[!htb]
   \includegraphics*[width=\columnwidth]{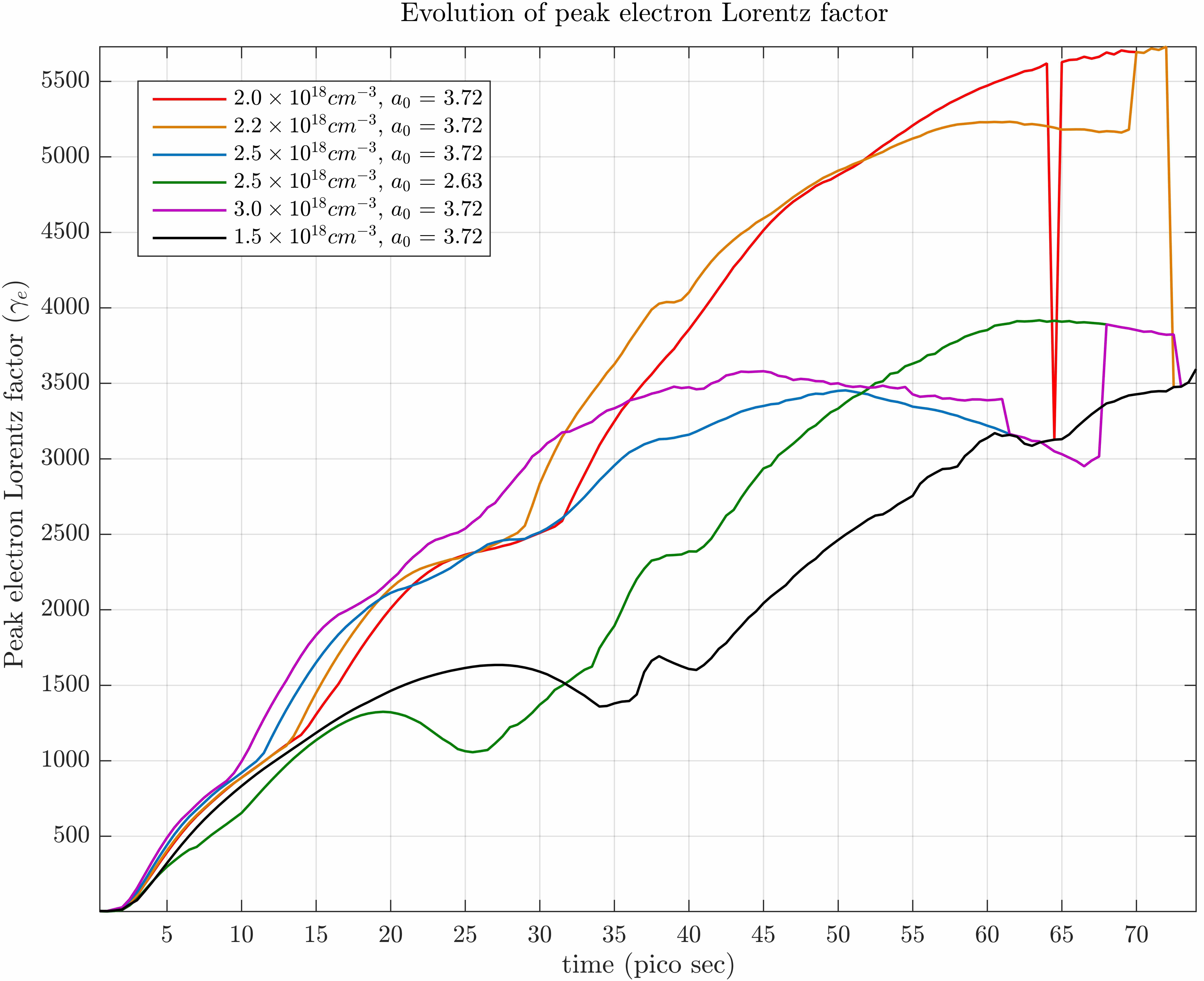}
   \caption{ 2D PIC simulations showing the evolution of the peak Lorentz factor for the f/40 focusing optics parameters for laser energy $\rm{\mathcal{E}_L}$ = 10J.}
   \label{Fig6:pic-beam-energy-evolution}
\end{figure}

We present the properties of the accelerated electrons that are injected by the unique {\it optical shock} driven asymmetric {\it elongating bubble}. Due to multiple injection events in the {\it strongly mismatched regime} the particle stream can only be loosely-termed as a beam with the \textsc{gemini} f/40 parameter set. Laser-plasma accelerated beam properties are dictated by the non-linearities of laser-plasma interactions. There is as strong dependence of plasma wake phase velocity and plasma wake dimensions on the instantaneous laser field amplitude which in turn depends upon the non-linear response of the plasma electrons. The matched regime which operates at an initially enforced equilibrium is considered most suitable for generating low energy spread beams with high consistency. In the strongly mismatched regime we see that the laser and the bubble properties critically depend upon the plasma density due to reasons explored above and thus expect the beam properties to reflect this.

In Fig.\ref{Fig6:pic-beam-energy-evolution} a unique property applicable in the realm of laser acceleration is plotted, it is the peak Lorentz factor of the accelerated electrons or indirectly the kinetic energy of the most energetic electrons. It is observed that the peak Lorentz factor clearly has two distinct phases which are represented in the two {\it different slopes} of the parabolic evolution of peak Lorentz factor in time. The parabolic shape is the characteristic of the linearly decreasing electric field as the beam gains energy over the back half of the bubble. The first slope (of the parabola in time) is observed until around 25ps, where the first set of injected particles have over-run the receding wake and thus de-phased from the accelerating phase. The second slope is observed after 25ps (after optical shock driven injection event around 24ps) where the peak Lorentz factor rises to $\gamma_p > 3000$ for $n_0 = 1.5 ~ - ~ 3.0 \times 10^{18} \rm{cm^{-3}}$ and $\gamma_p > 4000$ for $n_0 = 2 ~ \& ~ 2.2 \times 10^{18} \rm{cm^{-3}}$. This inflection point also corresponds to a large injection event triggered by the elongation of the bubble.

\begin{figure}[!htb]
   \includegraphics*[width=6.0in]{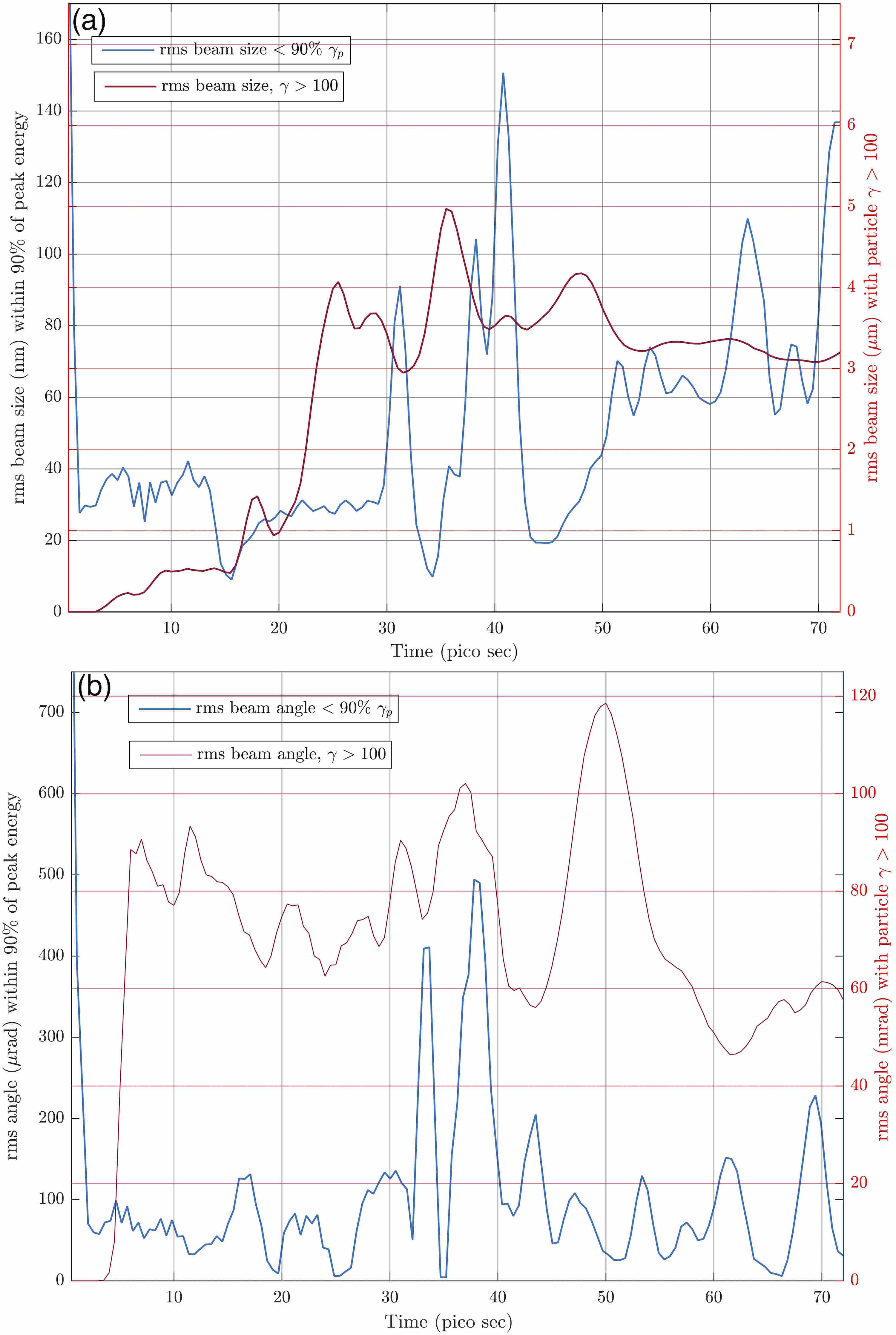}
   \caption{ Beam root-mean-square (rms) transverse-size (in a) and angular-size (in b) evolution with propagation distance at $n_0 = 2\times 10^{18}\rm{cm^{-3}}$ from 2D PIC simulations.}
   \label{Fig7:pic-beam-property-evolution}
\end{figure}

In Fig.\ref{Fig7:pic-beam-property-evolution} we show the root-mean-square beam transverse size (in a) and the root-mean square beam angle (in b) for $n_0 = 2 \times 10^{18} \rm{cm^{-3}}$ corresponding to the case for the highest kinetic energy electrons. There are two different metrics which are plotted for both these parameters. These two metrics are necessary because as the accelerated electrons are spread over a large volume of the full 6D phase-space it is hard to properly define $\sigma_r$ and $\sigma_{\theta}$ for laser accelerated electrons, unlike beams from conventional RF cavity based accelerators. The first metric we have adopted takes into account all the particle with $\gamma>100$ and second metric only looks at all the particles with $\gamma > 0.9\times\gamma_p$. So, the second metric is tuned towards understanding the properties of the highest energy particles.

We find that the rms-size in the y-direction for the component of the beam with the highest energy particles (within 90\% of $\gamma_p$, where $\gamma_p$ evolves as shown in Fig.\ref{Fig6:pic-beam-energy-evolution}) is around an average of 50 nm, through most of the evolution. However around the exit we observe that this size is closer to 100nm and this can be confirmed to be for the multi-GeV beam (as $\gamma_p \simeq 5500$ at $2\times 10^{18}\rm{cm^{-3}}$). The rms-size in the y-direction for all the particles with $\gamma>100$ is on average around 3.5 $\mu m$.

The rms-angle (based on $p_y/p_x$) for the component of the beam with the highest energy particles is around 100 $\mu \rm{rad}$. Whereas the rms-angle for the particles with $\gamma>100$ is on average around 80 mrad.

The effective geometrical emittance for the high-energy component of the beam can be estimated as $\epsilon_p = \text{rms-y}_p \times \text{rms-}\theta_p \simeq 10^{-5}$ mm-mrad. The corresponding normalized emittance $\epsilon_{p-n} = \gamma_p \times \epsilon_p = 0.04 \text{mm-mrad}$. The observation of distinct properties of the particles at the peak energy in comparison to all particle above 50 MeV points towards the {\it adiabatic damping} effect of the geometrical emittance of particles as they undergo acceleration in the plasma. It is also interesting to note the conservation of the emittance of the highest energy particles of the beam, which can be qualitatively seen by the anti-correlation between the $\text{rms-y}_p$ and $\text{rms-}\theta_p$ in Fig.\ref{Fig7:pic-beam-property-evolution}(a) and (b). Thus, the multi-GeV component of the beam behaves more like a conventional particle beam. As the beam exits the plasma, it enters in-homogeneous plasma which is likely to degrade its emittance primarily due to coherent effects and thus it is important to model the density structures closer to the plasma exit. 

\section{Conclusion}\label{sec:conclusion}
We have presented an analysis of multi-GeV acceleration in the strongly mis-matched regime which is conventionally considered sub-optimal for energy gain. It is shown that the adjusted-$a_0$ model based upon the physics of the self-guided laser pulse squeezing down to the matched spot-size and remaining close to it, is valid and justifiable. Further analysis correlates the surges in the laser electric field, due to radial envelope oscillations, to the triggering of optical shock state which understandably depends upon the initial density. Thus, it is shown that the mis-matched regime and the adjusted-$a_0$ model apply only when an optical shock state of the laser pulse is reached over a narrow range of densities. We have also shown that the optical shock excitation event leading to the strongest dis-balance between the longitudinal and the radial ponderomotive force causes a distinct inflection point in the peak Lorentz factor time evolution curve. This corresponds to a unique self-injection method resulting form the strong mismatch triggering the optical shock state which naturally couples the high-quality trapped beam to the large accelerating fields with longer de-phasing lengths in the elongated bubble. These results become more attractive for practical viability because it is experimentally known to be easy to controllably focus to a larger Gaussian focal spot. Thus launching larger focal spot laser pulses in the strongly mismatched regime may be a novel path towards high energy beams with large self-injected charge of high quality with much higher overall efficiency.

\section*{Acknowledgement}\label{sec:ack}
We acknowledge funding from STFC for the support of the John Adams Institute of Accelerator Science by grants ST/J002062/1 and ST/P000835/1. The EPOCH code used in this research was developed under UK Engineering and Physics Sciences Research Council grants EP/G054940/1, EP/G055165/1 and EP/G056803/1. All simulations were carried using the Imperial High Performance Computing systems.


\begin{thebibliography}{99}
\footnotesize
\setlength{\itemsep}{0pt}	

\bibitem{tajima-dawson}
Tajima, T., Dawson, J. M., 
	\textit{Laser Electron Accelerator}, 
	\href{http://link.aps.org/doi/10.1103/PhysRevLett.43.267}{Phys. Rev. Lett. {\bf 43}, pp.267-270 (1979)}, doi: 10.1103/PhysRevLett.43.267.

\bibitem{Max-1974}
Max, C. E., Arons, J., Langdon, A. B.,
	\textit{Self-Modulation and Self-Focusing of Electromagnetic Waves in Plasmas}
	\href{https://doi.org/10.1103/PhysRevLett.33.209}{Phys. Rev. Lett. 33, 209, (1974)}, doi: 10.1103/PhysRevLett.33.209.
	
\bibitem{Sun-1987}
Sun, G. Z., Ott, E., Lee, Y. C., Guzdar, P., 
	\textit{Self-focusing of short intense pulses in plasmas}, 
	\href{http://dx.doi.org/10.1063/1.866349}{Phys. of Fluids {\bf 30}, 526 (1987)}, doi: 10.1063/1.866349.
	
\bibitem{Sprangle-1987}
Sprangle, P., Cha-Mei Tang ; E. Esarey
	\textit{Relativistic Self-Focusing of Short-Pulse Radiation Beams in Plasmas},
	\href{http://dx.doi.org/10.1109/TPS.1987.4316677}{IEEE Transactions on Plasma Science, Vol. 15, Iss. 2, pp.145-153 (1987)}, doi: 10.1109/TPS.1987.4316677.
	
\bibitem{Esarey-1996}
E. Esarey, P. Sprangle, J. Krall, and A. Ting, 
	\textit{Overview of plasma-based accelerator concepts},
	\href{http://dx.doi.org/10.1109/27.509991}{IEEE Trans.Plasma Sci. 24, 252 (1996)}, doi: 10.1109/27.509991.
	
\bibitem{Pukhov-bubble-2002}
Pukhov, A., Meyer-ter-Vehn, J.,
	\textit{Laser wake field acceleration: the highly non-linear broken-wave regime}
	\href{http://dx.doi.org/10.1007/s003400200795}{J. Appl Phys B (2002) 74: 355.}, doi: 10.1007/s003400200795.
	
\bibitem{mangles-2004}
Mangles, S. P. D., Murphy, C. D., Najmudin, Z., Thomas, A. G. R., Collier, J. L., Dangor, A. E., Divall, E. J., Foster, P. S., Gallacher, J. G., Hooker, C. J., et. al.,
	\textit{Monoenergetic beams of relativistic electrons from intense laser-plasma interactions}
	\href{http://dx.doi.org/10.1038/nature02939}{Nature {\bf 431}, 535 (2004)}, doi:10.1038/nature02939.

\bibitem{Pukhov-2005}
Pukhov, A., Gordienko, S.,
	\textit{Scalings for ultrarelativistic laser plasmas and quasimonoenergetic electrons}
	\href{http://dx.doi.org/10.1063/1.1884126}{Physics of Plasmas 12, 043109 (2005)}, doi: 10.1063/1.1884126.

\bibitem{Lu_PRSTAB_2007}
Lu, W., Tzoufras, M., Joshi, C., Tsung, F. S., Mori, W. B., Vieira, J., Fonseca, R. A., Silva, L. O., 
	\textit{Generating multi-GeV electron bunches using single stage laser wakefield acceleration in a 3D nonlinear regime}
	\href{http://dx.doi.org/10.1103/PhysRevSTAB.10.061301}{Physical Review Special Topics - Accelerators and Beams 10, 061301 (2007)}, doi:10.1103/PhysRevSTAB.10.061301.
	
\bibitem{Kneip-2009}
Kneip, S., et. al.,
	\textit{Near-GeV Acceleration of Electrons by a Nonlinear Plasma Wave Driven by a Self-Guided Laser Pulse},
	\href{https://doi.org/10.1103/PhysRevLett.103.035002}{Phys. Rev. Lett. 103, 035002 (2009)}, doi:10.1103/PhysRevLett.103.035002.
	
\bibitem{chen-lens-1986}
Chen, P.,
	\textit{Plasma Focusing of High energy beams},
	\href{http://www.slac.stanford.edu/cgi-wrap/getdoc/slac-pub-4049.pdf}{SLAC-PUB-4049, August 1986}

\bibitem{optical-shock}
Gaeta, A. L.,
	\textit{Catastrophic Collapse of Ultrashort Pulses},
	\href{https://doi.org/10.1103/PhysRevLett.84.3582}{Phys. Rev. Lett. 84, 3582, (2000)}, doi:10.1103/PhysRevLett.84.3582.
	
\bibitem{sahai-thesis-2015}
Sahai, A. A., 
	\textit{On Certain Non-linear and Relativistic Effects in Plasma-based Particle Acceleration},
	\href{http://search.proquest.com/docview/1753637333}{Ph.D. dissertation, Duke university, July 2015}

\end{thebibliography}
\end{document}